# FINAL DESIGN OF THE PRODUCTION SSR1 CRYOMODULE FOR PIP-II PROJECT AT FERMILAB*

J. Bernardini†, V. Roger, D. Passarelli, M. Parise, G. Romanov, J. Helsper, M. Chen, M. Kramp, F. Lewis, B. Squires, T. Nicol, P. Neri [1]
FNAL, Batavia, IL 60510, USA, [1] University of Pisa, Pisa, 56126, Italy

## Abstract

This contribution reports the design of the production Single Spoke Resonator Type 1 Cryomodule (SSR1 CM) for the PIP-II project at Fermilab. The innovative design is based on a structure, the strongback, which supports the coldmass from the bottom, stays at room temperature during operations, and can slide longitudinally with respect to the vacuum vessel. The Fermilab style cryomodule developed for the prototype Single Spoke Resonator Type 1 (pSSR1), the prototype High Beta 650 MHz (pHB650), and preproduction Single Spoke Resonator Type 2 (ppSSR2) cryomodules is the baseline of the present design. The focus of this contribution is on the results of calculations and finite element analyses performed to optimize the critical components of the cryomodule: vacuum vessel, strongback, thermal shield, and magnetic shield.

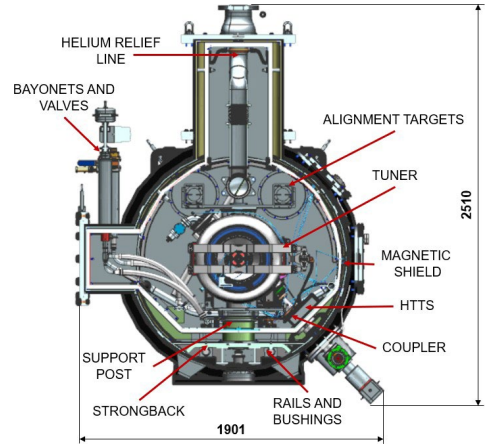

Figure 1: SSR1 CM transverse cross section showing the main components and subsystems (dimensions in mm).

## INTRODUCTION

The PIP-II linac [1] will utilize a total of two production SSR1 cryomodules (CMs) to accelerate H- ions from 10 MeV to 35 MeV. A prototype SSR1 CM has already been constructed, tested, and validated at Fermilab [4]. The valuable insights gained during testing have been incorporated into the design of the production CM. The design of the SSR1 CM is based on a groundbreaking concept developed at Fermilab known as the Fermilab style cryomodule [2, 3]. This design takes into consideration the standardization strategy established for the PIP-II CMs [?, 5]. To streamline the process of assembly, and minimize movement of the beamline components and ancillaries during the cooldown, a full-length strongback is utilized to support the coldmass and the beamline components. The strongback is designed to slide into the vacuum vessel during the CM assembly process, and to be maintained at room temperature during operations. A High Temperature Thermal Shield (HTTS) and Low Temperature Thermal Source (LTTS), along with connections for intercepts are made available between the inner surface of the vacuum vessel and the 2 K helium to reduce radiation and conduction heat transfers. The current PIP-II beam optics design requires that each SSR1 cryomodule contains eight identical SSR1 cavities and four focusing lenses with beam position monitors. Each cavity is equipped with one high-power RF coupler, and one tuner.

Cavities and focusing lenses are supported by individual support posts, which are mounted on the strongback, situated between the vacuum vessel and the HTTS. Alignment plates are incorporated between cavities/solenoids and the support posts, and they allow for 5 degrees of freedom (DOF) which are needed to align beamline components during the CM assembly process. In addition, the cavities and solenoids are equipped with reference targets that serve the purpose of monitoring their movements throughout various stages of the CM assembly, transportation, pump-down, and cooldown processes. These reference targets play a crucial role in ensuring the accurate tracking and assessment of any displacements or shifts that may occur [8, 9]. A two-phase helium pipe runs the length of the cryomodule and is connected to the cavities via Ti-SS transition joints. The focusing lenses are connected to the two-phase pipe using thermal straps. The two-phase pipe is linked to the relief line through the top hat and to the pumping line through the bayonets on the lateral extension of the vessel. The heat exchanger, as well as the interfaces with the 2 K relief line and pressure transducers, are positioned on the top hat of the vacuum vessel. Inside the vessel, an inner frame supports a global magnetic shield.

## MAIN CRYOMODULE COMPONENTS

### Vacuum Vessel

The vacuum vessel is constructed with a carbon steel (ASTM A-516) cylindrical shell. It is securely anchored to alignment stands, bolted to the floor, using bottom supports and features lugs for lifting purposes. The upstream and downstream ends of the vessel are sealed with endcaps, which are equipped with view-ports. The vessel shell includes various ports for functionalities such as input RF power couplers, access, instrumentation, vacuum pump-out, and safety relief. The total length of 5572 mm (620 mm shorter than SSR2) and an outer diameter of 1502.5 mm (same as SSR2). The primary function of the vacuum vessel extends beyond containment, as it also serves as a structural support for the strongback tray. The strongback tray can slide within the vessel through a system of rails and bushings, as depicted in 3. To ensure stability and secure positioning, the strongback tray can be locked to the vessel using a central lock-out stud, as illustrated in 2. This lock-out stud establishes a fixed point of reference with respect to which the strongback can contract and expand in response to temperature changes, such as those that may occur during a loss of vacuum. This design feature allows the strongback to accommodate thermal variations while maintaining its structural integrity.

* Work supported by Fermi Research Alliance, LLC under Contract No. DEAC02- 07CH11359 with the United States Department of Energy, Office of Science, Office of High Energy Physics.
† jbernard@fnal.gov

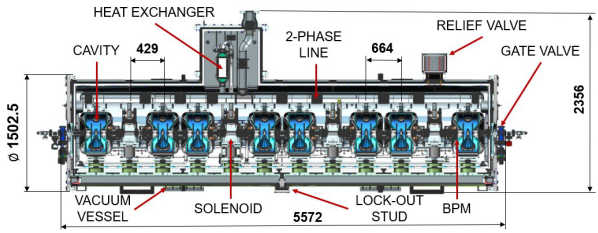

Figure 2: SSR1 CM longitudinal cross section showing main components and subsystems (dimensions in mm).

Finite element analyses (FEA) were conducted [10] to assess the impact of external atmospheric pressure, in the presence of insulating vacuum inside, on the mutual alignment of the cavities and solenoids. For this analysis, a vessel shell thickness of 11.2 mm was considered, slightly below the nominal thickness of 12 mm. The displacement of the beamline axis resulting from external pressure was evaluated at 12 specific locations along the beamline's axis, corresponding to the center of each cavity and solenoid. The results of these analyses are presented in 4. To minimize displacement of the beamline components and stresses in the welds connecting the top and side ports to the main shell, the number and placement of the vessel's stiffening rings were optimized. 4 displays the results achieved with two stiffening rings, which resulted in a maximum displacement of 0.1 mm in the 'x' direction. The permissible alignment errors for the system were determined to be 1 mm RMS for the cavities and 0.5 mm RMS for the solenoids. These alignment criteria were taken into account during the analysis and optimization process to ensure that the system meets the required standards.

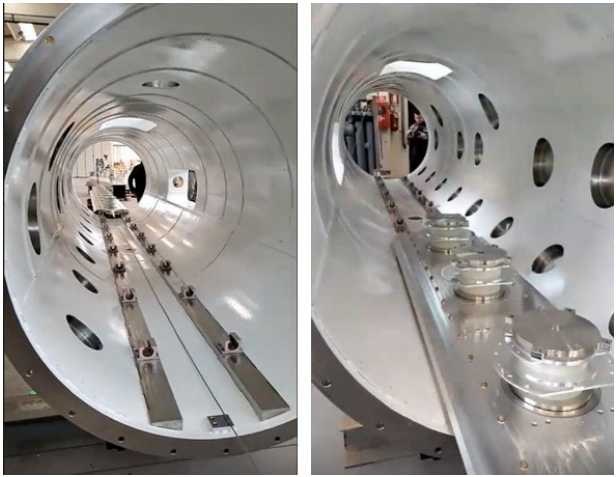

Figure 3: SSR2 insertion

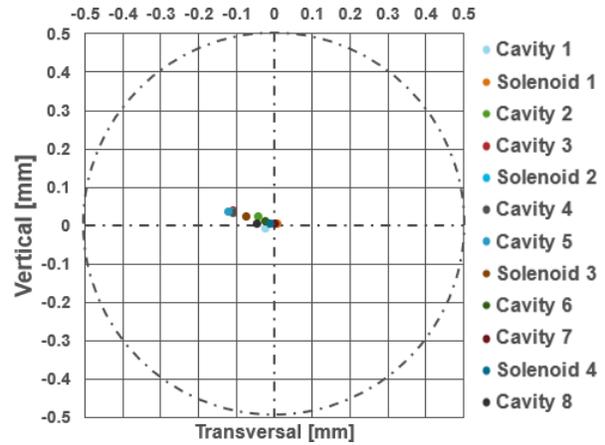

Figure 4: Cavities and solenoids transverse misalignment after pump down. The admissible misalignment is 0.5 mm RMS.

To ensure the structural integrity of the vacuum vessel design, an elastic stress analysis was conducted to evaluate its resistance against plastic collapse during lifting operations. The analysis revealed that the maximum vertical deformation experienced by the vessel during lifting is 1.1 mm. The highest equivalent stress was observed in the weldment of the lifting bracket, with each weld modeled with a volume derated by 55% to account for the minimum joint efficiency specified in Table UW-12 of ASME Section VIII, Division 1 [7]. The maximum equivalent stress resulting from the analysis was found to be 43 MPa, well below the maximum allowable stress specified in ASME Section II, Part D, which is 138 MPa. Buckling analysis was performed to predict instability of the vessel under external atmospheric pressure with vacuum inside. The first mode load multiplier is 23.8, which is higher than the minimum acceptable load multiplier of 2.5.

*Strongback*

The SSR1 strongback incorporates an Al-6061 T6 extrusion that is bolted to two parallel rails made of carbon steel. To enhance its structural stability, two Al-6061 T6 I-beams are added for reinforcement. It is worth noting that using the same material for both the extrusion and the I-beams helps mitigate stresses on the bolts resulting from temperature fluctuations. This lesson was learned from the design of the ppSSR2 strongback, where the I-beams were made out of SS 304. The strongback tray is designed to be smoothly inserted into the vacuum vessel by sliding the rails into open plain bearings, or bushings, that are securely bolted to the vessel. To prevent any undesired movement during handling, transportation, and operations, a central pin is incorporated to immobilize the strongback tray on the rails. In total, 16 bushings are strategically positioned within the vessel, with eight on the right and eight on the left side of beamline axis. The optimization of the number of bushings is aimed at facilitating the insertion of the coldmass into the vessel and ensuring the strongback's structural stiffness to meet the requirements for handling and transportation. The uniform number of bushings for both SSR1 and SSR2 CMs allows for the utilization of the same tooling during the coldmass insertion phase of the CM assembly.

The maximum rails' deflection, achieved when the last cav-ity on the downstream side of the coldmass is overhanging (just before the engagement of the tips of the rail into the bushings) is estimated to be 0.41 mm, as shown in Fig. 7.

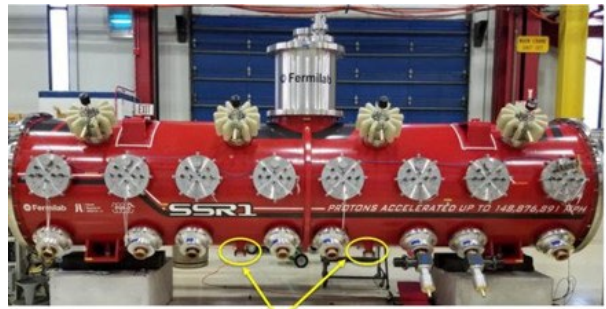

Figure 6: Heaters installed on the prototype SSR1 vacuum vessel to warm the strongback up.

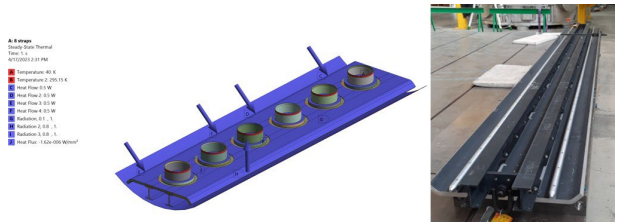

Figure 5: SSR1 strongback equilibrium temperature and high emissivity coating.

The bushings selected for this application have a self aligning feature that make the 0.41 mm deflection acceptable. Stresses in the strongback, rails, and stiffeners are well below the yield strengths of the materials.
To ensure the maintenance of alignment achieved during the string assembly phase throughout operation, it is imperative that the average equilibrium temperature of the strongback remains above $T_{avg} > 283$ K, with a maximum temperature differential $\Delta T < 5$ K. However, initial measurements indicated that the equilibrium temperature of the pSSR1 strongback was initially lower than the required threshold. To address this issue, a strategic approach was taken by placing heaters on the external surface of the vacuum vessel to warm the strongback. This enabled a sensitivity analysis to be conducted, which revealed that the emissivity of the strongback plays a significant role in determining the equilibrium temperature. As a result, it was decided to apply a high emissivity coating (expected emissivity $\epsilon = 0.8$) to the bottom surface of the production SSR1 strongback. Additionally, thermal straps will be installed at each coupler port location to establish a thermal connection between the vacuum vessel and the strongback. Thermal analyses were performed to evaluate the effect of the thermal straps and the high emissivity coating on the strongback's equilibrium temperature. Analyses' results are summarized in Table 1. The analyses were performed by setting the emissivity of the internal surface of the vessel $\epsilon = 0.1$ and by imposing a 1.62 W/m² heat flux from the strongback to the HTTS. The strongback temperature will be monitored continuously during CM testing with five temperature sensors. The movement of cavities and solenoids will be monitored continuously during the cooldown by using H-BCAMs [8,9].

Table 1: Strongback Equilibrium Temperature

| Boundary Conditions | $T_{avg}$ [K] | $\Delta T$ [K] |
|---|---|---|
| No Thermal Straps | 288.1 | 0.1 |
| 6 Thermal Straps | 289.4 | 0.24 |
| 8 Thermal Straps | 289.8 | 0.18 |

*Thermal Shield*

The High Temperature Thermal Shield (HTTS) is designed to minimize heat loads on the Low Temperature Thermal Source (LTTS) and the 2K volume. It achieves this by providing thermal intercepts (by means of thermal straps) and effectively preventing radiation between room temperature and 2 K components. The HTTS is constructed using specific materials: aluminum alloy Al 1100-H12 (3 mm and 6 mm thick for the upper and lower shields respctively) for the sheets and aluminum alloy Al 6061 T6 (6 mm thick) for the extrusion, which acts as a conduit for the helium gas. The HTTS structure is supported solely from the bottom, where it rests on aluminum rings that have been shrink fitted onto support posts. The shield has the ability to slide along the longitudinal direction on eleven posts, but it is fully constrained on only one post. The HTTS is convection cooled by helium gas flowing in the extrusion with a nominal inlet temperature $T_{in}$ = 40 K and pressure $P_{in}$ = 13 bar. The temperature differential across the HTTS in operations shall be less than 30 K and the temperature at the interface with current leads shall be less than 65 K. A variability of ±5 K is expected in the linac on the nominal He supply temperature. Moreover, the He inlet temperature differential between the 1st and 2nd cryomodule in the linac is expected to be lower than 1K. Therefore, a maximum helium inlet temperature of 46 K shall be expected for the production SSR1 CM.

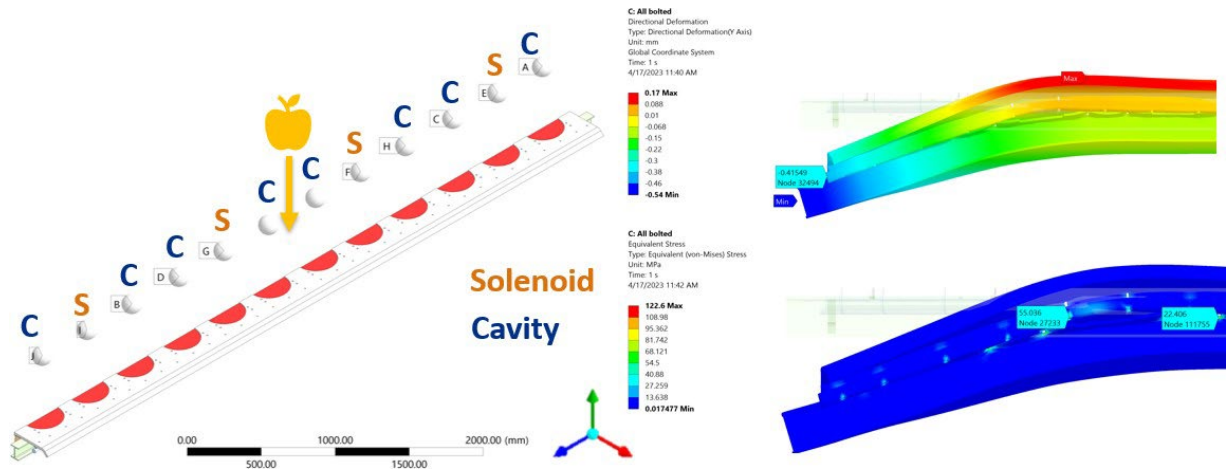

Figure 7: Strongback maximal static deflection during the insertion of the coldmass into the vessel.

FE thermal analysis shows that the minimum He mass flow rate that would allow having a temperature lower than 65 K at the interface with current leads is 7 g/s, as shown in Fig. 8. The resulting temperature differential across the shield is 20 K.

The resulting maximum shrinkage of the shield is

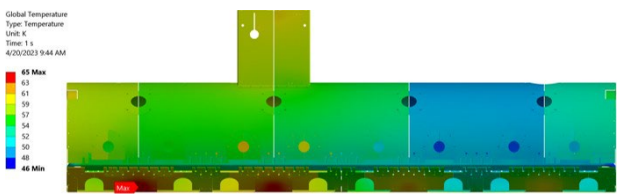

Figure 8: Thermal shield temperature distribution with a 46 K He supply temperature and 7 g/s mass flow rate.

14 mm, while the shrinkage at the interface with current leads is 11 mm. The HTTS will be cooled down at rate of 10 K/hour. Transient thermal analysis shows that the maximum temperature differential across the shield resulting from a 15 K/hour cooldown rate would be 109 K, which results in a maximum membrane plus bending stress in the fillet welds of 39 MPa. The resulting stress is well below the allowable limit of 130 MPa.

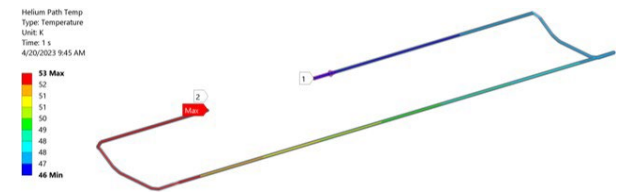

Figure 9: Helium temperature distribution with a 46 K He supply temperature and 7 g/s mass flow rate.

## Global Magnetic Shield

The SSR1 cryomodule incorporates a global magnetic shield that has been designed to attenuate the Earth's magnetic field to below 15 mG at the cavity surface.
To achieve this, the upper portion of the magnetic shield is bolted to a frame welded to the inner surface of the vacuum vessel, while the lower portion is bolted to the strongback. Unlike the magnetic shield frames used in previous designs such as pHB650 and ppSSR2, the SSR1 implementation utilizes longitudinal straight strips exclusively. This lesson learned from previous designs highlights that straight strips are easier to manufacture and can be welded with greater accuracy to the inner surface of the vessel. The upper and lower portions of the shield are meticulously closed with shunts wherever possible to minimize openings. Additionally, caps mounted in correspondence to the view ports further contribute to reducing the magnetic flux density on the cavity surface.

The magnetic shield material chosen for SSR1 is a 3 mm thick 80% Nickel-Iron alloy sheet, conforming to ASTM A753-85, Type 4. For simulation purposes, the magnetic permeability of the shield's material is set to $\mu$ = 40,000, while all components internal to the cryomodule are considered nonmagnetic. Through numerical optimization, the mag-netic shielding design has successfully achieved the desired goal by reducing the maximum magnetic flux density on the cavity surface to 9.7 mG, taking into account the effect of the carbon steel vacuum vessel. This result is illustrated in Fig. 11.

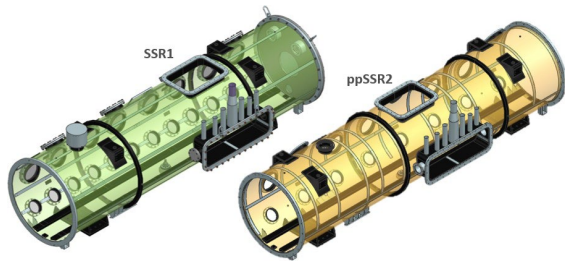

Figure 10: Comparison between SSR1 (left) and ppSSR2 (right) global magnetic shield frames.

After fitting the strongback and the magnetic shield into the vessel, measurements will be taken with a flux-gate type gaussmeter at the height of the portion of the cavities which is expected to see the highest amount of magnetic flux.

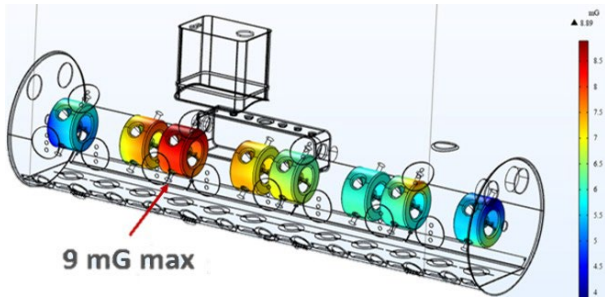

Figure 11: Magnetic flux density distribution on the surface of the cavities.

## CONCLUSION

Extensive finite element analysis and calculations were conducted to ensure that the critical components of the SSR1 CM adhere to the specified technical requirements. Specifically, the design of the vacuum vessel was carefully optimized to minimize misalignments of cavities and solenoids during the pump down process. The vacuum vessel design has been validated to ensure that it can withstand the lifting forces without plastic collapse and maintain stability under external pressure conditions. Structural and thermal analyses on the strongback tray were performed to verify the feasibility of the coldmass insertion and the strongback's temperature distribution in operation, which should allow maintaining the alignment of cavities and solenoids. The HTTS design was demonstrated to meet the technical requirements given the He supply temperature and mass flow rate that can be expected in the PIP-II linac. The incorporation of the global magnetic shield, along with the careful design considerations and optimizations, ensures that the SSR1 cryomodule effectively attenuates the Earth's magnetic field below the specified design requirements. The magnetic shield design allows having a maximal magnetic flux density of 9.7 mG on the surface of the cavities.